\documentclass[prl,onecolumn,aps,superscriptaddress]{revtex4-1}

\usepackage{bm}
\usepackage{times}
\usepackage[
            colorlinks,linkcolor=blue,
            citecolor=blue,
            pdfstartview=FitH,
            plainpages=false]
            {hyperref}
\usepackage{amsmath}
\usepackage{amssymb}
\usepackage{amsthm}
\usepackage{amsfonts}
\usepackage{times}
\usepackage{enumerate}
\usepackage{latexsym}
\usepackage{ifpdf}
\usepackage{graphicx}
\usepackage{color}
\usepackage{makeidx}
\usepackage{ulem}

\expandafter\ifx\csname package@font\endcsname\relax\else
\expandafter\expandafter
\expandafter\usepackage
\expandafter\expandafter
\expandafter{\csname package@font\endcsname}
\fi
\begin{document}

\title{Engineered Mott ground state in LaTiO$_{3+\delta}$/LaNiO$_3$ heterostructure}

\author{Yanwei Cao}
\email{yc003@uark.edu}
\affiliation{Department of Physics, University of Arkansas, Fayetteville, Arkansas 72701, USA}
\author{Xiaoran Liu}
\affiliation{Department of Physics, University of Arkansas, Fayetteville, Arkansas 72701, USA}
\author{M. Kareev}
\affiliation{Department of Physics, University of Arkansas, Fayetteville, Arkansas 72701, USA}
\author{D. Choudhury}
\affiliation{Department of Physics, University of Arkansas, Fayetteville, Arkansas 72701, USA}
\affiliation{Department of Physics, Indian Institute of Technology Kharagpur, Kharagpur 721302, India}
\author{S. Middey}
\affiliation{Department of Physics, University of Arkansas, Fayetteville, Arkansas 72701, USA}
\author{D. Meyers}
\affiliation{Department of Physics, University of Arkansas, Fayetteville, Arkansas 72701, USA}
\author{J.-W. Kim}
\affiliation{Advanced Photon Source, Argonne National Laboratory, Argonne, Illinois 60439, USA}
\author{P. J. Ryan}
\affiliation{Advanced Photon Source, Argonne National Laboratory, Argonne, Illinois 60439, USA}
\author{J. W. Freeland}
\affiliation{Advanced Photon Source, Argonne National Laboratory, Argonne, Illinois 60439, USA}
\author{J. Chakhalian}
\affiliation{Department of Physics, University of Arkansas, Fayetteville, Arkansas 72701, USA}

\begin{abstract}

In pursuit of creating cuprate-like electronic and orbital structures, artificial heterostructures based on LaNiO$_3$ have inspired a wealth of exciting experimental and theoretical results. However, to date there is a very limited experimental understanding of the electronic and orbital states emerging after interfacial charge-transfer and their connections to the modified band structure at the interface. Towards this goal,  we have synthesized a prototypical superlattice composed of correlated metal LaNiO$_3$ and doped Mott insulator LaTiO$_{3+\delta}$, and investigated its electronic structure by resonant X-ray absorption spectroscopy combined with X-ray photoemission spectroscopy, electrical transport and theory calculations. The heterostructure exhibits interfacial charge-transfer from Ti to Ni sites giving rise to an insulating ground state with orbital polarization and $e_\textrm{g}$ orbital band splitting. Our findings demonstrate how the control over charge at the interface can be effectively used to create exotic electronic, orbital and spin states.

\end{abstract}


\maketitle
Understanding and controlling the interactions between charge, spin, orbital, and structural degrees of freedom in transition-metal oxides (TMO) is at the center of modern condensed matter physics \cite{RMP-1998-Imada,Science-2000-Tokura,Khomskii-2014}. In recent years, inspired by the tremendously successful research on physics and applications of ultra-thin semiconductor heterostructures, multilayers of correlated complex oxides have become a platform of choice to generate the emergent electronic and magnetic states, unattainable in the bulk compounds \cite{RMP-2006-Ahn,NMat-2012-Hwang,RMP-2014-JC,NMat-2012-JC,Science-2010-Man,MRS-2013-Coey,ARMR-2007-Schlom,ARMR-2014-Stemmer,ARMR-2014-Bha,ARCMP-2011-Zubko,ARMR-2014-Ngai}. The inherent many-body nature of the correlated states, however, raises many fundamental questions which demand experimental validation and expansibility about the applicability of the concepts and formulas developed for semiconductor heterointerfaces \cite{RMP-2014-JC}. Recent experimental work on complex oxide interfaces \cite{NMat-2012-Hwang,RMP-2014-JC,NMat-2012-JC,Science-2010-Man,MRS-2013-Coey,ARMR-2007-Schlom,ARMR-2014-Stemmer,ARMR-2014-Bha,ARCMP-2011-Zubko,ARMR-2014-Ngai} revealed a remarkable importance of the electronic configurations of partially filled $d$-shell TM ions for understanding the emerging many-body phenomena  \cite{RMP-1998-Imada,Science-2000-Tokura,Khomskii-2014}, including the Ti 3$d^1$ configuration in the SrTiO$_3$-based two-dimensional electron gases \cite{NMat-2012-Hwang}, Cu 3$d^9\textrm{\underline {L}}$  configuration in the orbitally and magnetically reconstructed states at the manganate-cuprates interfaces (here $\textrm{\underline {L}}$ denotes a ligand hole on the oxygen ion) \cite{RMP-2014-JC, Science-2007-JC} and  Ni 3$d^8\textrm{\underline {L}}$ configuration in self-doped and orbitally polarized nickelate heterojunctions \cite{RMP-2014-JC,Khomskii-2014,NMat-2011-Ben,PRB-2011-Jian,PRL-2011-JC,EPL-2011-JF}. In order to induce a specific electronic configuration at the interface, charge-transfer (or electron doping) has been proven to be a particularly powerful tool to achieve this goal \cite{PRL-2013-Chen,PRL-2013-Chen2,PRL-2014-Disa}. Even more so than in the doped semiconductors \cite{RMP-1999-Stormer,Marle-2012, Gerter-2013,JVSTA-1990-Schubert}, charge-transfer with strong electron-electron correlations and frustrated spin and orbital interactions at the interface may give rise to unexpected collective quantum states not attainable with semiconductor heterojunctions \cite{NMat-2012-Hwang,RMP-2006-Ahn,RMP-2014-JC}.  
Understanding the mechanism of charge-redistribution between layers of Mott materials and the implications of a specific electron reconfiguration arising from the charge-transfer is therefore of a great necessity towards the rational design of applications based on strongly correlated electrons \cite{RMP-2014-JC,PRL-2013-Chen,PRL-2013-Chen2,PRL-2014-Disa}.

In semiconductor heterostrutures the charge-transfer can be successfully rationalized in terms of single-electron energy states to profile the energy band bending and the band alignments across the interface \cite{JVSTA-1990-Schubert,Sze-2007}.
Following this notion, we recap that within the class of complex oxides with 3$d$ electrons there are two types of Mott insulating behavior parameterized in the Zaanen-Sawatzky-Allen scheme by the relative magnitude of on-site Coulomb repulsion energy \textit{U}$_\textrm{dd}$ between $d$-shell electrons versus charge-transfer energy $\Delta_{\textrm{CT}}$ between oxygen $p$-shell and the TM $d$ state \cite{PRL-1985-Zaanen}; based on this  one can distinguish between Mott-Hubbard insulators (MHI, \textit{U}$_\textrm{dd}$ $<$ $\Delta_{\textrm{CT}}$) and charge-transfer insulators or charge-transfer metals (CTI or CTM, \textit{U}$_\textrm{dd}$ $>$ $\Delta_{\textrm{CT}}$) \cite{RMP-1998-Imada,Khomskii-2014}.  
To date the vast majority of experimental and theoretical work has been focused on charge doping at the interfaces between MHI and MHI \cite{PRB-2006-Lee,PRB-2013-Charle}, and MHI and normal metal \cite{PRL-2005-Oka,PRB-2007-Okamoto,PRB-2007-Yon,PRL-2015-Al}. 
Some of the most remarkable physical phenomena such as high-$T_\textrm{C}$ superconductivity and colossal magnetoresistance, however,  are observed in charge-transfer compounds characterized by the strong hybridization between oxygen 2$p$ and transition metal 3$d$ states, complex electronic configurations (e.g., mixing between $d^{n}$ and $d^{n}$$\textrm{\underline {L}}$ states) and small or even negative charge excitation gap $\Delta_{\textrm{CT}}$ \cite{PRL-1985-Zaanen,SSC-1997-Khomskii,RMP-1998-Imada,Khomskii-2014}. 
In these materials the role of the lower Hubbard band (LHB) is replaced by the oxygen states which in turn implies a very asymmetric physical character for the doped holes (mainly in oxygen levels) and doped electrons (mainly in transition metal \textit{d} levels),  e.g., correlated metal LaNiO$_3$ (LNO). With the original motivation to create a cuprate-like electronic and orbital structures \cite {PRL-2008-Cha,PRL-2009-Hans,RMP-2014-JC}, LNO-based perovskite heterostructures have attracted continuous interest \cite{NMat-2011-Ben,RMP-2014-JC,EPL-2011-JF,PRB-2013-Hoffman,PRL-2011-JC,PRL-2011-Kaiser,PRB-2011-Jian,APL-2010-Son,PRB-2009-May,PRB-2012-RS,PRL-2013-Chen,PRL-2013-Chen2,PRL-2011-Han,PRL-2014-Disa} in spite of the intriguing bulk properties of charge-transfer materials. However, experimentally very little is  known about the Mott carrier redistribution and their electronic reconfigurations at the heterointerface between MHI and CTM \cite{PRL-2013-Chen,PRL-2013-Chen2,PRL-2014-Disa}.

Towards this goal, we have synthesized and investigated a prototypical MHI/CTM heterostructure, [2u.c. LaTiO$_{3+\delta}$/ 2u.c. LaNiO$_{3}$]$\times$10 (2LTO/2LNO thereafter, u.c. = unit cells, $\delta$ $\sim$ 0.34 is the concentration of oxygen excess from the ideal Ti$^{3+}$ state). The resulting 2LTO/2LNO heterostructure exhibits an exotic Mott ground state. To quantify this phenomenon, we investigated the interfacial charge-transfer from Ti to Ni sites and the reconstruction of the electronic structure by resonant soft X-ray absorption spectroscopy (XAS) at Ti, Ni L$_{2,3}$-  and O K-edges combined with X-ray photoemission spectroscopy (XPS), electrical transport, and first-principles calculations. X-ray linear dichroism (XLD) spectroscopy was carried out to reveal the orbital polarization and unexpected Ni $e_\textrm{g}$ band splitting. Our findings highlight how the control over charge at the interface can be effectively used to create exotic electronic, orbital and spin states. 

\begin{figure*}[t!]
\vspace{-0pt}
\includegraphics[width=1\textwidth]{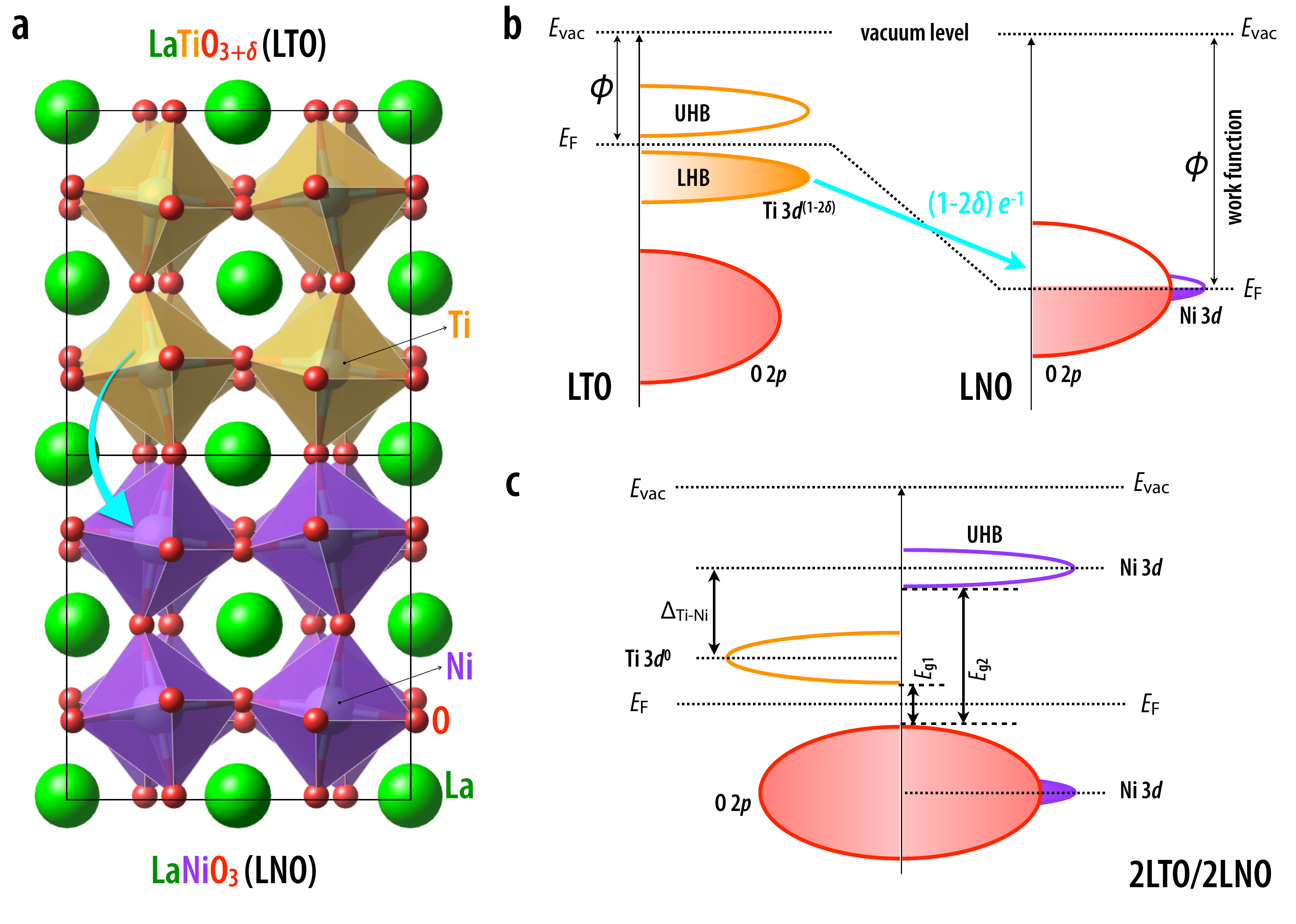}
\caption{\label{} \textbf{Crystal and band structures of 2LTO/2LNO.} (\textbf{a}) Schematic view of the atomic arrangement. (\textbf{b}) and (\textbf{c}) Sketch of the energy bands before (b) and after (c) the formation of 2LTO/2LNO interface. The cyan solid arrows in (a) and (b) indicate the  direction of interfacial charge-transfer (electron, $e^{-1}$) from Ti to Ni sites. Here, LHB (UHB) is the lower (upper) Hubbard band; $E_\textrm{g1}$ (or $E_\textrm{g2}$) is the charge gap between the highest occupied state (a mixture of oxygen 2$p$ and Ni 3$d$ states \cite{PRL-2013-Chen}) and the bottom of empty Ti 3$d$ (or Ni 3$d$) state, whereas $\Delta_{\textrm{Ti-Ni}}$ is the relative energy difference between empty Ti and Ni 3$d$ states.}
\end{figure*}

\textbf{Results}

\textbf{Interfacial charge-transfer.}
As shown in Fig.~1, driven by the difference between Fermi levels $E_\textrm{F}$ or chemical potentials in constituent layers across the junction 2LTO/2LNO (see Fig.~1a and Supplementary Fig.~1), in a conventional view the charge redistributes near the interface. As the components of 2LTO/2LNO superlattice, the electronic configuration of CTM LaNi$^{3+}$O$_3$ is a mixture of low-spin 3$d^7$ and high-spin 3$d^8$$\textrm{\underline {L}}$ states with the Fermi energy passing though the strongly mixed Ni-O valence states \cite{PRL-2011-Scherwitzl, PRB-1995-Gar}, whereas as an archetypal MHI ($\sim$~0.2~eV gap) bulk LaTi$^{3+}$O$_3$ has only one electron (3$d^{1}$) occupying the LHB \cite{NJP-2004-Mochizuki,PRB-1995-Okimoto} and its Fermi energy level $E_\textrm{F}$ is much higher than that of LNO ($\sim$ 2 eV difference) \cite {AFM-2012-Greiner}, as schematically shown in Fig.~1b. By aligning the interfacial bands with respect to the continuing oxygen $p$ states  on either side of the interface \cite{PRL-2013-Chen,PRL-2014-Kle}, the resulting Ti 3$d$ band energy position becomes significantly higher than the Fermi energy of the LNO; this in turn implies a one-way charge redistribution from the Ti 3$d$ band of LTO into the partially filled Ni $d$ and O $p$ states of LNO. Recent density functional theory (DFT + $U$) calculations \cite{PRL-2013-Chen} further tested this naive picture and suggested a full electron charge-transfer, i.e., Ti $d^{1}$ + Ni $d^{7}$~$\rightarrow$~Ti $d^{0}$ + Ni $d^{8}$ (Fig.~1c). On the other hand, since the experimental electronic configuration of LNO is a mixture of Ni $d^7$ and d$^8$$\textrm{\underline L}$ states, the charge-transfer may also result in the appearance of additional interfacial electronic states, i.e., Ti $d^{1}$ + Ni $d^{8}$$\textrm{\underline L}$~$\rightarrow$~Ti $d^{0}$ + Ni $d^{8}$ and Ti $d^{1}$ + Ni $d^{8}$$\textrm{\underline L}$~$\rightarrow$~Ti $d^{0}$ + Ni $d^{9}$$\textrm{\underline L}$.

To investigate the experimental veracity of the theory, we measured the electronic structures of Ti and Ni to track the charge-transfer by element-specific XAS in total fluorescence yield mode (TFY with the bulk probing depth) and by in-situ XPS. 
As seen in Fig.~2a, the features of the Ti L$_{2,3}$-edge in the 2LTO/2LNO sample show excellent agreement with the Ti$^{4+}$ charge and are markedly  different from the spectra of Ti$^{3+}$.
This result provides a strong evidence for the occurrence of the charge-transfer Ti $d^{(1-2\delta)}$~$\rightarrow$ Ti $d^{0}$ and implies that almost all  of the $t_\textrm{2g}$ electrons from Ti sites are transferred elsewhere. The flow of the charge is further verified by the complimentary XAS measurements  at the Ni L$_{2,3}$-edge, which clearly shows a strong increase of the Ni charge state, i.e., Ni $d^{7}$~$\rightarrow$~Ni $d^{(8-2\delta)}$ (see Fig.~2b). A comparison to the bulk reference spectra of Ni$^{2+}$ (double peaks at $\sim$ 870.2 eV and 871.2 eV) and Ni$^{3+}$ (single main peak at $\sim$ 871.6 eV) attests that in the 2LTO/2LNO superlattice the Ni final state is indeed  a mixture of Ni$^{2+}$/Ni$^{3+}$ (double peaks), which is also affirmed by the calculated XAS lineshape dependence on the Ni electronic configuration (see Supplementary Fig.~2). To further corroborate these findings, the interfacial charge-transfer phenomenon was studied by measuring the core-level electronic structures of Ti and Ni with in-situ XPS (see Supplementary Fig.~3); as determined by XPS the resulting charge states of Ni  and Ti in the 2LTO/2LNO sample are in excellent agreement with those obtained by XAS at Ti L$_{2,3}$- and Ni L$_{2,3}$-edges. Moreover, as revealed by the angle-dependent (ex-situ) XPS (see  Supplementary Fig.~4), the pronounced Ni$^{2+}$ peak near the interface of metallic 2LTO/8LNO (see Supplementary Fig.~5) further confirmed the interfacial charge-transfer from Ti to Ni sites.

\begin{figure*}[t!]
\vspace{-0pt}
\includegraphics[width=1\textwidth]{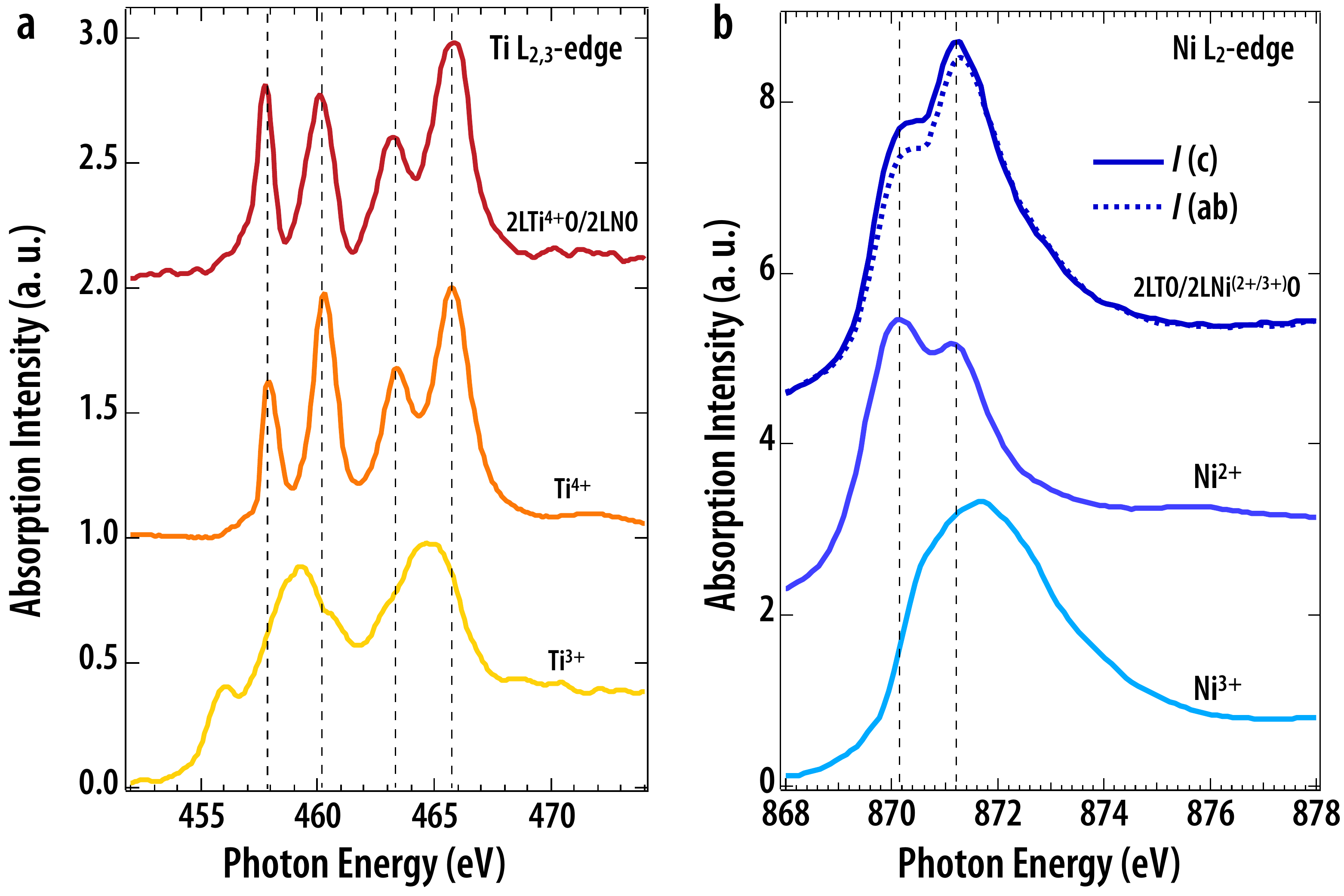}
\caption{\label{} \textbf{X-ray absorption spectroscopy of 2LTO/2LNO.} (\textbf{a}) Ti L$_{2,3}$-edge. The reference spectra for Ti$^{4+}$ and Ti$^{3+}$ were measured on a SrTi$^{4+}$O$_3$ single crystal and YTi$^{3+}$O$_3$ film ($\sim$ 100 nm on TbScO$_3$ substrate \cite{APL-2013-Misha}), respectively. (\textbf{b}) Ni L$_{2,3}$-edge. The reference samples are bulk Ni$^{2+}$O and  LaNi$^{3+}$O$_3$. Out-of-plane [$I$(c), dark blue solid line, $E\parallel$ c and $E$ is the linear polarization vector of the photon] and in-plane [$I$(ab), dark blue dashed line, $E\parallel$ ab] linearly polarized X-ray were used to measure XAS of 2LTO/2LNO at Ni L$_{2,3}$-edge. Black dashed lines are guidelines for peak positions. All spectra were collected and repeated more tan two times with bulk-sensitive total fluorescence yield (TFY) detection mode at room temperature.}
\end{figure*}

\begin{figure*}[t!]
\vspace{-0pt}
\includegraphics[width=1\textwidth]{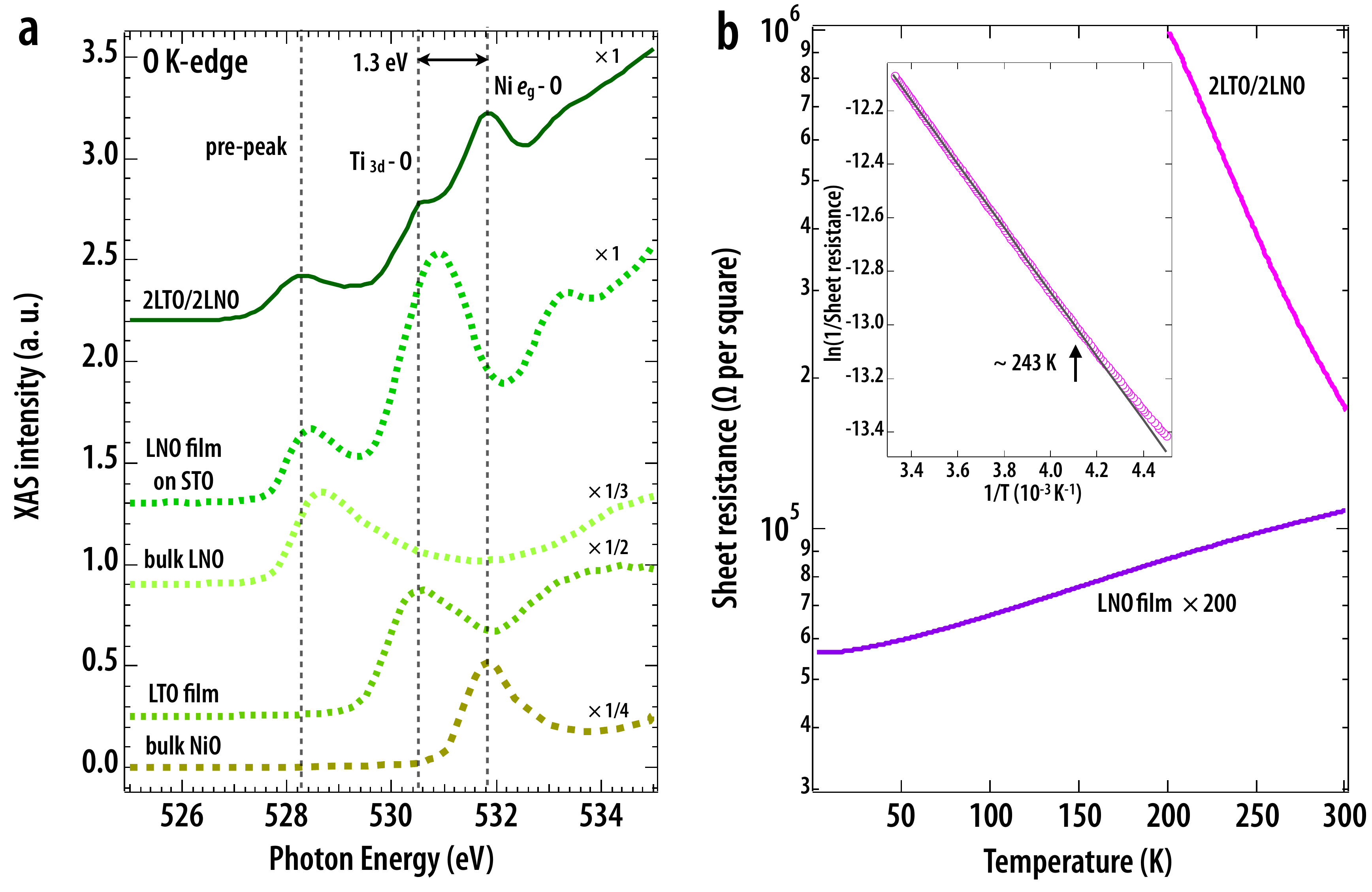}
\caption{\label{} \textbf{Oxygen K-edge spectra and electrical transport of 2LTO/2LNO.} (\textbf{a}) Normalized XAS spectra at O K-edge. The black dashed lines indicate the assignments of three key features: the pre-peak ($\sim$ 528.5 eV) and hybridized Ti 3$d$-O ($\sim$ 530.5 eV) and Ni $e_\textrm{g}$-O ($\sim$ 531.8 eV) states by comparing with the spectra of four reference samples LaNi$^{3+}$O$_3$ film (10 u.c. on SrTiO$_3$ substrate), bulk LaNi$^{3+}$O$_3$, LaTi$^{3+}$O$_3$ film (20 u.c. on TbScO$_3$ substrate \cite{APL-2013-Misha}) and bulk Ni$^{2+}$O (the data of NiO was adapted from Ref. \cite{NCom-2014-Lin}). (\textbf{b}) Temperature-dependent sheet resistances of the SL 2LTO/2LNO and the reference LaNiO$_{3}$ film (20 u.c.). Note, the sheet resistance of LNO film is $\times 200$. Inset: resulting fit of the conductance of 2LTO/2LNO (black solid line) yielding an activation gap $E_\textrm{g1}\sim 0.2 \pm 0.01$ eV \cite{APL-2012-Sri}. }
\end{figure*}

\textbf{Electronic reconstruction.} With the confirmed large interfacial charge-transfer from Ti to Ni sites, an important question arises: how does the interfacial charge-transfer alter the fundamental physical properties (i.e., electronic configuration and band structure) of the 2LTO/2LNO SL? 
First, we discussed the emergent electronic configuration. As mentioned above, the channels of interfacial charge-transfer Ti $d^{1}$ + Ni $d^{7}$~$\rightarrow$~Ti $d^{0}$ + Ni $d^{8}$ and Ti $d^{1}$ + Ni $d^{8}$$\textrm{\underline L}$~$\rightarrow$~Ti $d^{0}$ + Ni $d^{8}$ are both open at the interface. 
Experimentally, due to the strong  hybridization between Ni 3$d$-states and oxygen 2$p$-states at the Fermi level, XAS at O $K$-edge becomes another important way  to  probe the charge states of Ni and Ti mixed with ligand holes. 
As seen in Fig.~3a and Supplementary Fig.~6, in 2LTO/2LNO  the oxygen K-edge spectra clearly show a characteristic low energy pre-peak at $\sim$ 528.5 eV, which arises from the ligand holes \cite{PRL-2011-JC}. In sharp  contrast, a direct comparison  to the LTO and LNO reference samples immediately shows that the pre-peak at the 2LTO/2LNO interface is strongly suppressed due to the filling oxygen ligand holes with the transferred electrons from Ti sites (also see Supplementary Fig.~6). 
Based on the absence of the pre-peak feature in the LaTi$^{3+}$O$_3$ and  Ni$^{2+}$O reference samples, these data imply that the strong suppression of the pre-peak intensity results from the filling of holes on oxygen by the interfacial charge-transfer into the Ni \textit{d}-band. As a result, this process induces the formation of the $d^{8}$ state and the strong suppression of the $d^8$$\textrm{\underline L}$ configuration.  Also, we point out at the expected difference between the theory (full charge-transfer of one electron) \cite{PRL-2013-Chen} and the experimental observation of less than one electron transfer, i.e., (1-2$\delta$) electron;  the observed deviation from the full charge-transfer in theory  is due to the reduced electron filling from unity, 2$\delta$ of the Ti $d$-band. Combined with the mixed $d^7$ and $d^8\textrm{\underline L}$ ground state of bulk LNO this factor results in the observed peculiar electronic configuration of $d^8$, $d^7$, and $d^8\textrm{\underline L}$ states  that appears at the interfacial NiO$_2$ layer in 2LTO/2LNO.

\begin{figure*}[t!]
\includegraphics[width=0.8\textwidth]{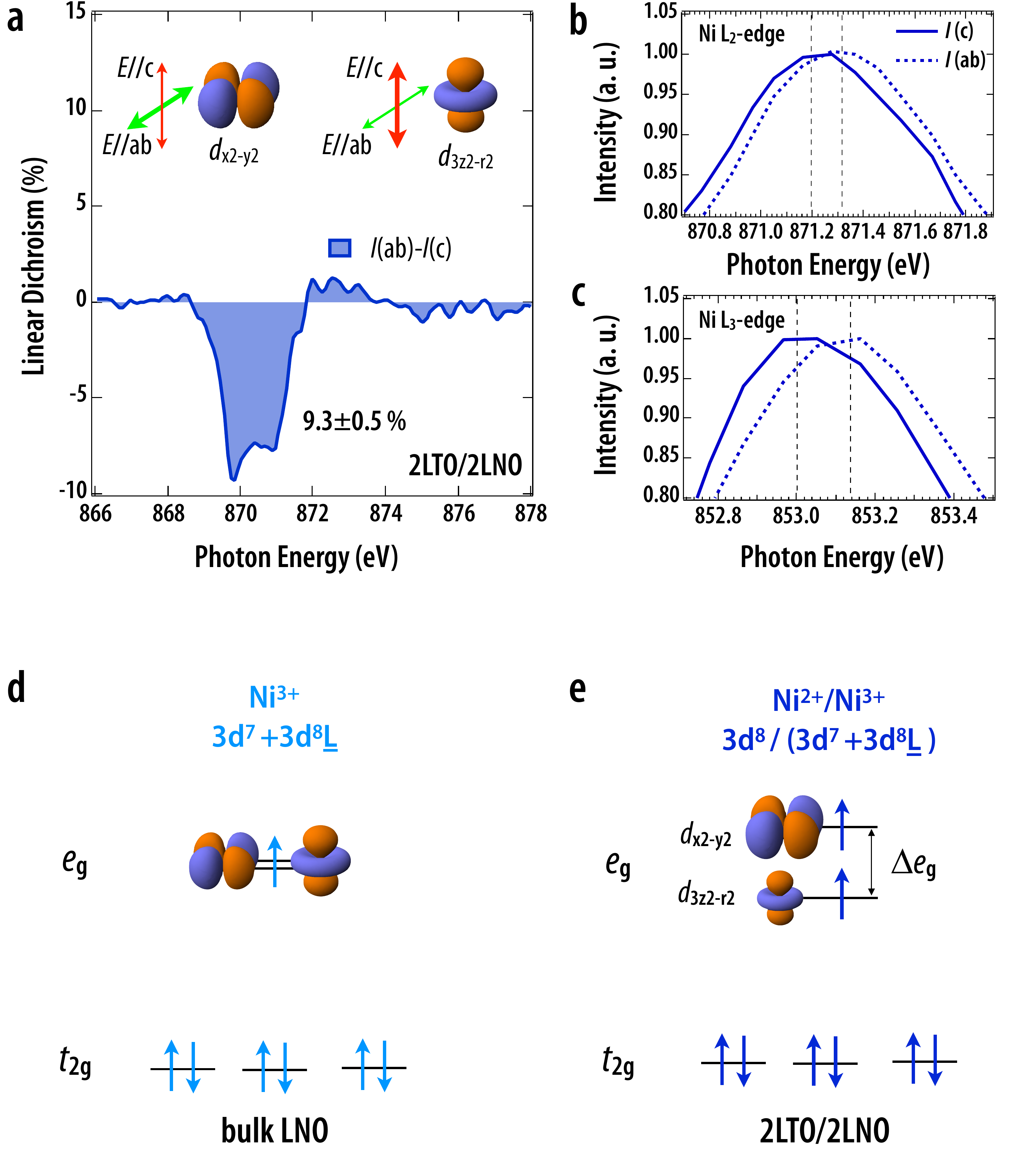}
\caption{\label{} \textbf{Linear dichroism and reconstructed states.} (\textbf{a}) XLD [$I$(ab)-$I$(c)] of 2LTO/2LNO interface (extracted from Fig.~2b). Green (in-plane) and red (out-of-plane) arrows label the direction of linear polarization vector ($E$) of the photon. (\textbf{b}) and  (\textbf{c}) XAS of 2LTO/2LNO at Ni L$_{2,3}$-edge showing the Ni $e_\textrm{g}$ band splitting ($\sim$~0.12 for L$_2$ and 0.15 eV for L$_3$, respectively) of $d_{3\textrm{z}^2-\textrm{r}^2}$ (lower) and $d_{\textrm{x}^2-\textrm{y}^2}$ (higher) orbitals. Black dashed lines are guidelines for peak center positions. (\textbf{d}) and (\textbf{e}) Sketch of the engineered electronic, orbital and spin states via interfacial charge-transfer for bulk LNO \cite{PRL-2011-JC} and SL 2LTO/2LNO, respectively. In SL 2LTO/2LNO it forms peculiar electronic configuration (Ni 3$d^7$, 3$d^8$ and 3$d^8\textrm{\underline L}$), orbital polarization ($n_{\textrm{d}_{\textrm{x}^2-\textrm{y}^2}}$ $>$ $n_{\textrm{d}_{3\textrm{z}^2-\textrm{r}^2}}$) and $e_\textrm{g}$ band splitting ($\triangle$ $e_\textrm{g}$ $\sim$ 0.15 eV). Blue arrows indicate the spin configurations of Ni sites.}
\end{figure*}

Next, we discuss the reconstructed band structure at the interface.
Because of the observed strong reconstruction of electronic configuration, it is natural  to  anticipate  a similarly strong modification of the band structure near the interface.
As predicted by the theory \cite{PRL-2013-Chen} and illustrated in Fig.~1c, two opening gaps $E_\textrm{g1}$ and $\Delta_{\textrm{Ti-Ni}}$ are expected to appear at the interface. First we estimated a magnitude of the charge gap $E_\textrm{g1}$ by measuring the temperature-dependent electrical transport properties of the 2LTO/2LNO and LNO reference films. As immediately seen in Fig.~3b, the LNO thin film grown  at the same conditions as the SL shows a metallic bulk-like behavior ($\sim$~280 $\Omega$ per square at 2 K) from room temperature down to 2~K. In sharp contrast, the SL 2LTO/2LNO displays a highly insulating behavior with a very large sheet resistance increasing from $\sim$~175 k$\Omega$ per square at 300~K to $\sim$ 1 M$\Omega$ per square at 200~K  exceeding the measurement range of the transport setup. This insulating behavior of the 2LTO/2LNO implies the charge excitations gap opening in 2LTO/2LNO. The resulting fit to the transport data shown in inset of Fig.~3b yields a value of  $E_\textrm{g1}$ $\sim$ 0.20 eV  $\pm$ 0.01 eV; this is in a accord with the theoretical prediction of the $\sim$ 0.4 eV charge-transfer gap \cite{PRL-2013-Chen}. 

Next, we estimate the value of the gap $\Delta_{\textrm{Ti-Ni}}$ between empty Ti $t_\textrm{2g}$ and Ni $e_\textrm{g}$ bands by measuring XAS of the 2LTO/2LNO film at O K-edge. 
In a simple ionic model, the configuration of oxygen is O 1$s^2$2$s^2$2$p^6$ and thus the transition 1$s$~$\rightarrow$~2$p$ is blocked in the absorption process due to it is a fully occupied 2$p$-shell for the O ion. 
In real materials, however, due to the strong hybridization, the covalent bonding between the transition metal ion and oxygen can introduce a sizable spectral weight of oxygen 2$p$ character in the total unoccupied density of states  \cite{PRB-1989-Groot,PRB-2007-Wu,JPCM-2013-Mos, JMCC-2013-Cho}.
As a result, O K-edge XAS provides a complimentary way to probe the relative energy position of the TM ion.
We also  point out, that compared to the L$_{2,3}$-edge XAS reflecting the absorption process for the specific TM ion,  O K-edge  provides a convenient way to measure the relative energy position of the unoccupied bands of both TM ions (Ti and Ni)  present  in the 2LTO/2LNO\ heterostructure  \cite{JMCC-2013-Cho}. 
A direct comparison to  the reference samples allows to assign the two peaks at  $\sim$ 530.5 eV and $\sim$ 531.8 eV shown in Fig.~3a to  the hybridized oxygen 2$p$ with Ti 3$d$ and Ni $e_\textrm{g}$ bands, respectively, and then to extract the value of  the Mott gap $\Delta_{\textrm{Ti-Ni}}$ $\sim$ 1.3 eV. 
Assuming that the bandwidth of Ti 3$d$-O and Ni $e_\textrm{g}$-O bands is roughly the same (see Fig.~3a), and with the known value of $E_\textrm{g1}$ $\sim$ 0.2 eV  the estimated value  of the  correlated  gap  $E_\textrm{g2}=(E_\textrm{g1}+\Delta_{\textrm{Ti-Ni}}$) is $\sim$ 1.5 eV; this value is in a remarkable agreement with the theoretically predicted value  of $\sim$ 1.5 eV \cite{PRL-2013-Chen}. The above observation of the two gaps-opening in the excitation spectrum  lends strong support to the notion of a strong modification of the band structure at the interface triggered by the redistribution of correlated charges. It is also noted  some additional contributions (e.g., disorder effect, electron-electron interactions, and charge/spin order) \cite{PRL-2011-Scherwitzl,NJP-2011-Moon,NNano-2014-King} may be involved in to the enhanced carrier localization of 2LTO/2LNO. However, their contributions are not dominant comparing with the large $\sim$ 1.5 eV correlated gap of 2LTO/2LNO. As demonstrated by the angle-dependent XPS on 2LTO/8LNO (see Supplementary Fig.~4), the formation of Ni$^{2+}$ accompanied with the strongly suppressed density of states (DOS) near the Fermi energy level is primarily driven by the interfacial charge-transfer from Ti to Ni sites.

\textbf{Orbital reconstruction.}  With  the established strongly  altered $d$-band filling on Ni  and Ti  we investigated the orbital properties of these engineered states on Ni sites. To this end, the orbital polarization has been measured by XLD \cite{PRL-2011-JC,EPL-2011-JF,NMat-2011-Ben,PRB-2013-Wu} on several LNO-based hetrostructures and the result for the  LTO/LNO SL\  is shown in Fig.~4a. 
Based on the measured electronic state of Ni, one can anticipate that contribution to the XLD signal on Ni L$_{2}$-edge arises largely from the unoccupied Ni $d_{\textrm{x}^2-\textrm{y}^2}$ ($I$(ab)) and $d_{3\textrm{z}^2-\textrm{r}^2}$ ($I$(c)) states. As illustrated in Fig.~4a, those orbital  configurations can be probed  with in-plane ($E\parallel$ ab,  $E$ is the polarization vector of the photon) and out-of-plane ($E\parallel$ c) linearly polarized photons, respectively. 
In a good agreement with this expectation, the XLD spectra shown in Fig.~4a shows an ample degree of  orbital  polarization of $\sim$ 9.3\% at L$_{2}$-edge with the $d$-electron occupancy $n_{\textrm{d}_{\textrm{x}^2-\textrm{y}^2}}$ $>$ $n_{\textrm{d}_{3\textrm{z}^2-\textrm{r}^2}}$consistent with the first principle calculation prediction of $\sim$ 9\% (see Calculation details in Methods).

The surprising feature of the XLD data for 2LTO/2LNO is the presence of $e_\textrm{g}$ band splitting (Fig.~4b,c) that was not observed in tensile-strained ultra-thin LNO films or SL (1u.c.)LaNiO$_3$/(1u.c.)LaAlO$_3$  \cite{PRL-2011-JC, EPL-2011-JF}.
Note that the band splitting is generally estimated by the peak energy shift of XAS (between $I$(ab) and $I$(c)) with linearly polarized photons and the lineshape of XLD [$I$(ab)-$I$(c)] with multiple peak features does not infer the size of the band splitting directly.
In the  case of 2LTO/2LNO, however,  small tensile strain of  +1.04 \% causes the sizable $e_\textrm{g}$ band splitting.   
Specifically, as seen in inset of Fig.~4b,c, a direct inspection of the energy position for in-plane ($\sim$ 853.15 eV) and out-of-plane ($\sim$ 853.0 eV) absorption curves reveal that the out-of-plane absorption is $\sim$ 0.15 eV (0.12 eV) lower in energy than the in-plane absorption at Ni L$_3$(L$_2$)-edge. The difference implies the $e_\textrm{g}$ band splitting $\triangle e_\textrm{g}$ $\sim$ 0.15 eV between the states with Ni $d_{\textrm{x}^2-\textrm{y}^2}$ and $d_{3\textrm{z}^2-\textrm{r}^2}$ orbital character, as schematically illustrated in Fig.~4d,e. 
This observation lends strong support to the recently predicted by theory (see Calculation details in Methods) both band splitting and orbital polarization arise from the structural distortions at the interface. The apical oxygen atom of out-of-plane Ni-O-Ti bond approaches Ti and leaves away from Ni atoms whereas the in-plane Ni-O bond length is changed little and smaller than the out-of-plane Ni-O bond length. This extended out-of-plane Ni-O bond leads to the lowering of Ni $d_{3\textrm{z}^2-\textrm{r}^2}$ central band energy, on the other hand, it also obviously weakens the hybridization between O $2p$ and Ni $d_{3\textrm{z}^2-\textrm{r}^2}$ bands with suppressed virtual electron hopping. Therefore the larger hybridization between O $2p$ and Ni $d_{\textrm{x}^2-\textrm{y}^2}$ bands results in a higher electron occupancy at higher Ni $d_{\textrm{x}^2-\textrm{y}^2}$(minority spin) orbital band $n_{d_{\textrm{x}^2-\textrm{y}^2}}$ $>$ $n_{d_{3\textrm{z}^2-\textrm{r}^2}}$ which is very unusual. 

In conclusion, by synthesizing the 2LTO/2LNO interface as a prototypical system, we investigated the reconstruction of the local electronic structure on Ni and Ti at the interface between a doped Mott-Hubbard insulator and a charge-transfer metal. Our findings reveal  the occurrence of large charge-transfer from the Ti to Ni sites across the interface that results in the unusual electronic configurations of Ni 3$d$-electrons and lead to the strong modification of the band structure in the vicinity of the interface.
In addition, the XLD data show the presence of the large orbital polarization and energy splitting of the Ni $e_\textrm{g}$ band at the vicinity of  the interface characteristic of the Jahn-Teller distortion absent in either bulk rare-earth nickelates or other ultra-thin LNO based heterojunctions. 
We anticipate that these results will pave the way for follow up theoretical and experimental work with other important classes of charge-transfer interfaces to establish a discovery platform for exotic many-body quantum phenomena. 

\textbf{Methods}

\textbf{Experiment details.} High-quality superlattices (SLs) [2u.c. LaTiO$_{3+\delta}$/$n$u.c. LaNiO$_3$] $\times$ 10 ($n$ = 2 and 8, 2LTO/2LNO and 2LTO/8LNO thereafter, u.c. = unit cells) and reference samples were epitaxially grown by pulsed laser deposition (PLD) on 5 $\times$ 5 $\times$ 0.5 mm$^3$ (001)-oriented single crystal substrates (LaAlO$_{3}$)$_{0.3}$-(Sr$_{2}$AlTaO$_{6}$)$_{0.7}$ (LSAT, cubic, $a$~=~3.87 \AA), using a KrF excimer laser operating at $\lambda$~=~248~nm and 2 Hz pulse rate with 2~J cm$^{-2}$ fluence. The layer-by-layer growth was monitored by in-situ high pressure reflection-high-energy-electron-diffraction (RHEED) (see Supplementary Fig.~1). To match the growth conditions for both LaTiO$_{3+\delta}$ and LaNiO$_{3}$, the SLs 2LTO/2LNO and 2LTO/8LNO were grown under oxygen pressure $\sim$~50 mTorr and the temperature of the substrates was held at $580\,^{\circ}{\rm C}$ during the growth. After growth, all samples were cooled at about $15\,^{\circ}{\rm C}$ per minute rate to room temperature keeping oxygen pressure constant. A Mg anode was used for in-situ XPS measurements with double-pass cylindrical mirror analyzers (STAIB) at room temperature whereas for ex-situ angle-dependent XPS measurements with a hemispherical electron analyzer an Al anode with monochromator (PHI VersaProbe II) was applied. The sheet-resistances of the films were measured in van-der-Pauw geometry by Physical Properties Measurement System (PPMS, Quantum Design) from 300 to 2~K. XAS/XLD with total fluorescence yield (TFY) mode and X-ray diffraction (XRD) measurements (room temperature) were carried out at the 4-ID-C and 6-ID-B beamlines, respectively, of the Advanced Photon Source (APS, Argonne National Laboratory).

\textbf{Calculation details.} Calculated XAS at Ni L$_{2,3}$- and O K-edges of rhombohedral (R-3cH space group) LaNiO$_3$ \cite{JBTorrance1992} were carried out with the finite difference method near-edge structure (FDMNES) code \cite{YJoly2009}. In FDMNES calculations we used the full-multiplet-scattering (Green function) mode with a large cluster radius of 6 \AA~around the absorbing Ni atom. The XAS calculations were performed for various Ni 3$d^x \textrm{\underline L}$ (7 $<$ $x$ $\leqslant$ 8) configurations of LaNiO$_3$ (see Supplementary Fig.~2). To confirm the consistency of the calculated XAS spectra, we also performed XAS calculations using the multi-electron time-dependent density functional theory (TD DFT+$U$) with an on-site Coulomb energy on Ni of 6 eV. On the other hand, in the calculation of SL 2LaTiO$_3$/2LaNiO$_3$ by the DTF+$U$ method ($U_\textrm{Ni}$ = 6 eV and $U_\textrm{Ti}$ = 4 eV), the orbital polarization of unoccupied states is [(2-$n_{\textrm{d}_{\textrm{x}^2-\textrm{y}^2}}$)-(2-$n_{\textrm{d}_{3\textrm{z}^2-\textrm{r}^2}}$)]$/$[(2-$n_{\textrm{d}_{\textrm{x}^2-\textrm{y}^2}}$)+(2-$n_{\textrm{d}_{3\textrm{z}^2-\textrm{r}^2}}$)] $\sim$ -9\%, where $n_{\textrm{d}_{\textrm{x}^2-\textrm{y}^2}}$ and $n_{\textrm{d}_{3\textrm{z}^2-\textrm{r}^2}}$ are total electron occupancy (spin up plus spin down) with the $d_{\textrm{x}^2-\textrm{y}^2}$ and $d_{3\textrm{z}^2-\textrm{r}^2}$ orbital characters of the $e_\textrm{g}$ band, respectively; note the central energy position of $d_{\textrm{x}^2-\textrm{y}^2}$ band is higher than that of the $d_{3\textrm{z}^2-\textrm{r}^2}$ band (personal communication with A. J. Millis \& H. Chen).

\textbf{Acknowledgements}

The authors deeply acknowledges numerous fruitful discussions with Andrew Millis and Hanghui Chen.  J. C. and X. L. were supported by the Department of Energy grant DE-SC0012375 for synchrotron work at  the Advanced
Photon Source and material synthesis.
D. M. was primarily supported by the Gordon and Betty Moore Foundation EPiQS Initiative through Grant No. GBMF4534.  Y. C. and S. M. were supported by the DOD-ARO under Grant No. 0402-17291. This research used resources of the Advanced Photon Source, a U.S. Department of Energy (DOE) Office of Science User Facility operated for the DOE Office of Science by Argonne National Laboratory under Contract No. DE-AC02-06CH11357.

\textbf{Contributions}

Y. C. and J. C. designed the experiments. Y. C., X. L. and S. M. measured the electrical transport. Y. C. and M. K. collected the XPS data. Y. C., X. L., S. M., D. M., J. K. and P. R. measured the XRD. J. W. F. acquired the XAS data. D. C. and J. C. carried out the theoretical calculations. M. K. and Y .C. prepared the samples. Y. C. and J. C. analyzed the data. All authors discussed the results. Y. C., D. M., D. C., and J. C. wrote the manuscript with input from all authors. This work was supervised by J. C.

\section{Competing financial interests}
The authors declare there are no competing financial interests.

\clearpage
\newpage

\end{document}